# A Scalable Nanogenerator based on Self-Poled Piezoelectric Polymer Nanowires with High Energy Conversion Efficiency

*Richard A. Whiter, Vijay Narayan and Sohini Kar-Narayan\**

R. A. Whiter and Dr. S. Kar-Narayan
Department of Materials Science and Metallurgy, University of Cambridge, Cambridge, CB3 0FS, UK.
Email: sk568@cam.ac.uk

Dr. Vijay Narayan
Department of Physics, Cavendish Laboratory, University of Cambridge, Cambridge, CB3 0HE, UK.

Nanogenerators based on piezoelectric materials convert ever-present mechanical vibrations into electrical power for energetically autonomous wireless and electronic devices. Nanowires of piezoelectric polymers are particularly attractive for harvesting mechanical energy in this way, as they are flexible, lightweight and sensitive to small vibrations. Previous studies have focused exclusively on nanowires grown by electrospinning, but this involves complex equipment, and high voltages of ~10 kV that electrically pole the nanowires and thus render them piezoelectric. Here we demonstrate that nanowires of poly(vinylidene fluoride-trifluoroethylene) [P(VDF-TrFE)], grown using a simple and cost-effective template-wetting technique, can be successfully exploited in nanogenerators without poling. A typical nanogenerator comprising ~$10^{10}$ highly crystalline, self-poled aligned nanowires spanning ~2 cm$^2$ is shown to produce a peak output voltage of 3 V at 5.5 nA in response to low-level vibrations. The mechanical-to-electrical conversion efficiency of 11% exhibited by our template-grown nanowires is comparable with the best previously reported values. Our work therefore offers a scalable means of achieving high-performance nanogenerators for the next generation of self-powered electronics.

Energy harvesting from ambient vibrations has generated significant interest[1,2] as it offers a fundamental energy solution for small-power applications including, but not limited to, ubiquitous wireless sensor nodes, portable and wearable electronics, biomedical implants and structural/environmental monitoring devices. Nanoscale piezoelectric harvesters, also known as nanogenerators, convert small-scale vibrations into electrical energy thus offering a means of superseding batteries that require constant replacing/recharging, and that do not scale easily with the size of the device. Nanogenerators were first demonstrated using ZnO nanowires[3-5] and subsequent nanogenerators based on GaN nanowires,[6] $BaTiO_3$ nanowires, [7] $PbZr_xTi_{1-x}O_3$ nanowires[8] and nanoribbons[9,10] and poly(vinylidene fluoride) (PVDF) nanofibres[11] have all revealed promising energy harvesting performance. There has since been an ongoing concerted effort in developing this relatively new research field, connecting nanotechnology with the field of energy.[12]

Piezoelectric nanowires are particularly attractive for energy harvesting due to their robust mechanical properties and high sensitivity to typically small ambient vibrations.[13] The implications of these properties, in fact, go beyond energy harvesting, as nanowire-based nanogenerators have recently been shown to function as bio mechanical sensors,[14] sensitive pressure sensors[15] and precision accelerometers.[16] The challenge lies in the large scale production of low cost piezoelectric nanowires that can offer reproducible and reliable energy harvesting and/or sensor performance.

The polymer PVDF [(CH2-CF2)$_n$] exhibits good piezoelectric and mechanical properties with excellent chemical stability and resilient weathering characteristics.[17,18] PVDF thin films are thus commonly used as sensors and actuators.[19] However, the piezoelectric performance of PVDF is dependent on the nature of the crystalline phase present. Typically, PVDF occurs in the α, β and γ crystalline phases[20-22] and needs to be electrically poled (using an electric field of the order of 100 MV m$^{-1}$) and/or mechanically stretched[20,22] to achieve the polar β-phase that shows the strongest piezoelectric behavior.[21] P(VDF-TrFE) [(CH$_2$-CF$_2$)$_n$-(CHF-CF$_2$)$_m$] is a co-polymer that crystallizes more easily into the β-phase due to steric factors,[22] an advantage that we exploit in this work. Nanowires of PVDF and its copolymers have been previously incorporated into piezoelectric nanogenerators[11] but the relatively complex electrospinning fabrication process employed requires high voltages (5-50 kV) and specialized equipment. The associated high electric fields and stretching forces result in poled nanowires, however this fabrication process often suffers from poor control over nanowire size-distribution and alignment, and is yet to be conveniently and cost-effectively scaled up.[23] Here we report the growth of aligned P(VDF-TrFE) nanowires with a narrow size distribution using a simple, cost-effective and easily scalable template-wetting method,[24,25] where the template-induced space confinement promotes high crystallinity and preferential orientation of the lamellar crystals in the polymer nanowires.[26,27] This results in the enhancement of piezoelectric properties, even without the need for electrical poling. A nanogenerator fabricated using template-grown, self-poled P(VDF-TrFE) nanowires is shown have excellent electrical output when subjected to periodic vibrations. Using a circuit comprising a rectifier to convert its AC output to DC, and a bank of capacitors to store the harvested energy, the nanogenerator is shown to be capable of lighting a commercial light emitting diode (LED).

Aligned P(VDF-TrFE) nanowires were grown by the infiltration of a solution of the polymer into commercially available nanoporous anodized aluminium oxide (AAO) membranes of nominal pore diameter 200 nm and length 60 μm. Subsequent controlled evaporation of the solvent resulted in the formation of nanowires having diameter 196 ± 18 nm within the pores as confirmed by scanning electron microscopy (SEM) (see **Figure 1**a-e). Following the deposition and evaporation of the solvent, a ~10 μm-thick P(VDF-TrFE) film remained on the surface of the AAO template via which infiltration had been initiated. Our fabrication process was modified to completely remove this residual film such that the resulting template contained only nanowires. (See Supplementary Information S1). The vertical alignment of the P(VDF-TrFE) nanowires was enforced by the pores of the AAO template, resulting in a better alignment than in similar nanowires fabricated by electrospinning.[11] Our template-wetting process typically yielded 2.8 mg of P(VDF-TrFE) nanowires (equivalent to ~ $2 \times 10^8$ fibers) per hour for each AAO template (see Supplementary Information S2), and can thus be easily scaled up as the nanowires are readily processed from solution and easily deposited over large numbers of templates.

The AAO template can be dissolved to free the nanowires into solution if required (see Supplementary Information S3). X-ray diffraction (XRD) patterns of the freed P(VDF-TrFE) nanowires dispersed on a zero-diffraction silicon plate revealed peaks (**Figure 2**a) corresponding only to the (200) and (110) planes of the β crystalline phase.[28,29] In comparison the XRD pattern of P(VDF-TrFE) powder of the same composition showed peaks associated with α, β and γ phases indicating a mixture of crystallinities as expected for a powder specimen. The comparatively smaller β peak observed in the nanowire sample arises due to the low signal-to-noise ratio resulting from only sparse coverage of the silicon plate by the freed P(VDF-TrFE) nanowires.

Differential scanning calorimetry (DSC) and infrared (IR) spectroscopy studies were used to further characterize the P(VDF-TrFE) nanowires. The nanowire specimens for these studies were prepared by grinding P(VDF-TrFE) nanowire-filled AAO templates, and the results were compared with P(VDF-TrFE) powder specimen. The positions of the peaks in the DSC scans of the nanowire specimen (Figure 2b) corresponding to the ferroelectric-to-paraelectric transition at ~100°C and the melting transition at 148°C were found to be consistent with previous studies.[30-32] Comparison with the powder specimen revealed an increase in the relative height of the ferroelectric-to-paraelectric transition peak in the nanowire specimen and a shift of the peak to higher temperatures, indicating a higher proportion of the desired β crystalline phase in the nanowires.[31,32] This is further confirmed through IR spectroscopy studies. It was previously shown in electrically poled and/or annealed P(VDF-TrFE) films that an increased depth of the ~840 cm$^{-1}$ trough relative to the ~880 cm$^{-1}$ trough indicated an increased proportion of the β crystalline phase.[33-38] We observe a similar trend in the IR spectrum of our nanowire specimen, but not in the powder specimen (see Figure 2c), suggesting stronger β-phase character of the nanowires. This is consistent with recent studies which show that a higher proportion of the β crystalline phase in template-grown P(VDF-TrFE) nanowires and preferential orientation of the crystal lamellae occur solely due to the confinement effect of the nanoporous template.[26,27,38,39]

A nanogenerator, based on template-grown P(VDF-TrFE) nanowires, was fabricated by sputtering platinum electrodes (diameter $D$ ~1.7 cm) on either side of the nanowire-filled template, attaching thin copper wires for electrical access, and encapsulating the device in ~10 μm-thick poly(dimethylsiloxane) (PDMS) for protection (see **Figure 3**a). The nanogenerator comprising ~$10^{10}$ aligned nanowires was then impacted parallel to the length of the nanowires by the oscillating arm of a permanent magnet shaker in a bespoke energy harvesting measurement setup, as shown in the schematic in Figure 3b. (See Supplementary Information S4 for details). To generate maximum compressive normal force upon impact, the nanogenerator was rigidly fixed at the mean position of the oscillating arm. Figures 3c and 3d show typical output voltage and current characteristics respectively of the nanogenerator when impacted with low-frequency vibrations of amplitude $X = 1$ mm at frequency $f = 5$ Hz. The AC output signal at the same frequency arises as a result of the arm first compressing the device upon impact, causing a current to flow, followed by relaxation as the arm moves away giving rise to a current in the opposite direction. We observe a peak output voltage of 3 V and peak output current of 5.5 nA when the nanowires are subjected to a strain rate of 0.1% s$^{-1}$ (see Supplementary Information S5). The energy conversion efficiency of the template-grown P(VDF-TrFE) nanowires was found to be ~11% as estimated from the ratio between the maximum electrical energy output per cycle and the strain energy in the nanowires per cycle[3,40] (see Supplementary Information S6). This value is comparable to some of the highest reported efficiency values estimated in the same manner for nanogenerators based on electrospun nanowires,[40] and can be attributed to our template-grown nanowires having very few defects due to the observed high degree of crystallinity and confinement-induced preferential crystal lamellar orientation. Importantly, continuous impacting over 12 hours (~ 10$^6$ cycles) revealed no significant changes in the electrical output (see Figure 3e), and the results were found to be reproducible across several nanogenerators that were fabricated and fatigue-tested over several days. Cross-sectional SEM images of a nanogenerator taken before and after fatigue testing (Figures 3f & g) show no significant difference in the morphology of the device.

The energy harvesting performance of our template-grown nanowires can be explained by a simple model based on fundamental piezoelectric theory which gives the peak generated current from the nanowires as $i = d_{33}EA\dot{\varepsilon}$,[40] where $d_{33}$ is the piezoelectric charge constant, $E$ is the Young's modulus of the nanowires, $A$ is the total cross-sectional area, and $\dot{\varepsilon}$ is the applied strain rate. Considering that our device underwent compressive strain for a quarter of the oscillation period, $\dot{\varepsilon}$ is taken to be $4f\varepsilon$, where $\varepsilon$ is the strain. Substituting for $\varepsilon = \sigma/E$, where the stress generated $\sigma = V/g_{33}L$, $V$ being the output voltage, $g_{33}$ the piezoelectric coefficient and $L$ the length of the P(VDF-TrFE) nanowires, $i = 4fVd_{33}A/Lg_{33}$. Here, $A = 0.5\ \pi(D/2)^2$ as the AAO templates typically have 50% porosity. Taking $d_{33} = 35$ pC N$^{-1}$, $g_{33} = 0.43$ Vm N$^{-1}$,[41] $i$ is calculated to be 9.2 nA which is remarkably close to our observed value. The slight mismatch can be attributed to a fraction of the pores of the AAO template being only partially filled during the growth of the nanowires, and/or some of the P(VDF-TrFE) nanowires within the pores making poor contact with the platinum electrodes.

An electrical circuit was assembled to rectify the output AC signal and store the harvested energy using capacitors (see **Figure 4**a). An LED was connected to determine as part of a proof-of-principle study whether a single template-grown P(VDF-TrFE) nanogenerator was capable of lighting the LED following a charging period. The capacitors were first charged by impacting the nanogenerator for ~45 minutes ($X = 1$ mm, $f = 20$ Hz). The capacitors were then switched from parallel to series configuration (switch position 1 to 2 in Figure 4a) in order to step up the voltage, as highlighted in the inset of Figure 4b. The voltage then decreases until the LED switch is closed resulting in a rapid discharge across the LED (Figure 4b). Figure 4c shows photographs of the LED being lit during the discharge. The charging period of the capacitors can be significantly reduced by using a stack of nanogenerators connected in series electrically. A stack of three individual nanogenerators when impacted under the same conditions as above, was found to be capable of fully charging the capacitor bank in ~12 minutes thus reducing the time taken to light the LED (see Supplementary Information S7). The output current density per unit area for our stack of three nanogenerators was found to be 0.02 µA cm$^{-2}$ and can be further enhanced by adding more nanogenerators to the stack.

The overall efficiency of a single nanogenerator device comprising a P(VDF-TrFE) nanowire-filled AAO template encapsulated in PDMS was calculated as the ratio of electrical energy stored in the capacitors (Figure 4a) to the total mechanical work done in exciting the nanowires within the device in order to generate this electrical energy. The total mechanical work is the product of the number of cycles required to charge the capacitors and the work done in each cycle.[10] This yielded an overall device efficiency of 0.2% (see Supplementary Information S8) which can be attributed to a large proportion of the mechanical energy being lost to the AAO template and PDMS, leaving only a small fraction of mechanical energy actually transferred to the embedded P(VDF-TrFE) nanowires. This is also reflected in the observed slow temporal response of the nanogenerator seen in Figure 3d which arises due to the added presence of the template and PDMS. The overall efficiency of the nanogenerator can be improved by increasing the porosity of the templates, decreasing the thickness of the PDMS, and/or decreasing the Young's modulus of the template. Furthermore, nanogenerator designs involving freed template-grown P(VDF-TrFE) nanowires are expected to benefit from the inherently high energy conversion efficiency of the nanowires.

In conclusion, we have demonstrated working, high-output piezoelectric nanogenerators based on cost-effective and scalable template-grown nanowires of P(VDF-TrFE). The confined geometry of the templates promotes the oriented β crystalline phase required for superior piezoelectric energy harvesting performance of the nanowires, without the need for electrical poling. Our near room-temperature fabrication process is attractive as it can be easily implemented over large areas, and the resulting high-quality piezoelectric nanowires do not require high voltages for growth and/or poling. The as-grown, self-poled nanowires with high energy conversion efficiency are shown to be capable of converting low-level vibrations into sufficient electrical energy to light an LED in a scalable manner. Our results suggest that the nanogenerator described here can also be adapted to a variety of low-cost strain sensors with applications ranging from biomedicine to robotics. This work thus represents a crucial step in developing commercially feasible energy and/or sensor solutions for smart micro/nano-electronic devices.

**Experimental Section**

*Fabrication of nanowires:* P(VDF-TrFE) in powder form with a composition of 70:30 by weight of VDF:TrFE (Piezotech, France) was dissolved in butan-2-ol (Sigma-Aldrich) to a concentration of 10% by weight. The solution was sonicated for ∼20 minutes to ensure complete dissolution of the P(VDF-TrFE) powder. A few drops of the polymer solution were deposited using a pipette onto nanoporous AAO templates (Anapore, Whatman) of diameter 2 cm mounted on glass slides. The templates were then left for ∼24 hours on a hot plate at 60°C to allow for complete infiltration and evaporation, resulting in the formation of nanowires within the pores. A residual polymer film of thickness ∼10 μm was found to remain on the surface of the AAO templates onto which the polymer solution was dropped. Our fabrication process was thus modified by pre-coating the AAO templates with a thin ~15 nm layer of silver (see Supplementary Information S1 for details) to aid in the subsequent chemical and plasma etch process required to completely remove the residual polymer film.

*XRD measurements:* A Bruker D8 θ/θ diffractometer was used to study P(VDF-TrFE) powder and P(VDF-TrFE) nanowires drop cast onto a zero-diffraction silicon plate. To release the nanowires into solution, the AAO templates were submerged in vials with a minimal amount of 10% phosphoric acid solution and sonicated for ~1 hour, and then left submerged for ~24 hours to ensure complete dissolution of the templates. The vials were then topped up with deionised water and placed in a centrifuge (Sigma 1-14 Microfuge) for ~1 hour at 14,000 RPM. This resulted in the nanowires collecting at the bottom of the vial and allowed for the phosphoric acid solution to be removed and replaced with water. This process was repeated several times to wash out all of the phosphoric acid leaving freed nanowires suspended in water.

*DSC studies:* A differential scanning calorimeter (TA Instruments Q2000) was used to study the heat changes during the ferroelectric-to-paraelectric phase transition and melting of P(VDF-TrFE) powder and ground-up P(VDF-TrFE)-nanowire-filled AAO templates. A pestle and mortar were used for the grinding and ~3 mg of the specimen powders were weighed into thermally identical aluminium pans and sealed with aluminium lids before being loaded in to the calorimeter for measurement. Prior to measurement equilibrium was established at 20°C and heating rates of 10°C per minute were used for temperature ranges up to 200°C.

*IR spectroscopy studies:* An infrared spectrometer (Bruker Tensor 27) was used with an attenuated total reflectance (ATR)-FTIR attachment to study P(VDF-TrFE) powder, ground-up P(VDF-TrFE)-nanowire-filled AAO templates, as well as ground-up empty AAO templates. Following a baseline measurement, the specimen powders were placed on an IR-transparent high-refractive index crystal on which the IR light from the spectrometer is incident. Pressure was then applied using a small metal disk to press the specimen powders firmly against the crystal, and IR spectra were subsequently generated for each specimen.

*Fabrication of nanogenerator:* Platinum electrodes of thickness ∼100 nm were sputter-coated on either side of the P(VDF-TrFE)-nanowire-filled AAO templates. A shadow mask was used to prevent platinum from being coated completely to the edge of the template, with a few mm being left around the edge to prevent the possibility of shorting the top and bottom electrodes. Thin copper wires were then attached to each electrode using silver paint. For protection, the device was then encapsulated in 10 μm-thick polydimethylsiloxane (PDMS) via spin-coating[42] on either side.

*Energy harvesting setup:* A purpose-built steel frame was constructed to house a permanent magnet shaker (LDS Systems V100) to which an aluminium disk was attached, capable of impacting a nanogenerator held rigidly in the steel frame. (A top-down photo of this is shown in Supplementary Information S3). A dedicated amplifier (LDS Systems PA25E-CE) driven by a signal generator (Thurlby Thandor TG1304) was attached to the shaker to generate impacting oscillations of desired frequencies and amplitudes. The impacting arm underwent periodic oscillations at frequency $f$, given by $X \cos(\omega t)$, where $\omega = 2\pi f$ and $X$ is the maximum displacement. This amplitude of oscillations, $X$, was determined by measuring the maximum acceleration, $a$, of the impacting arm using an accelerometer (Brüel and Kjær DeltaTron 4517) stuck on to it, where $a = X \omega^2$. The copper wires from the electrodes of the nanogenerator being impacted were soldered into contact pads from which a data acquisition module (Keithley KUSB-3116) was used to record the electrical data on a computer (see schematic in Figure 3a) by being connected either directly to the contact pads or via a picoammeter (Keithley 6487) for output voltage or current measurements respectively. The electrical circuit shown in Figure 4a comprised a full-wave diode bridge rectifier consisting of 4 diodes (Taiwan Semiconductor 1N4148 A0) used to rectify the AC output of the nanogenerator, and a bank of 8 capacitors (Panasonic ECA1JM220) to store the harvested energy.

**Supporting Information**
Supporting Information is available from the Wiley Online Library or from the author.

**Acknowledgements**
We thank Dave Ritchie, Sam Crossley and David Muñoz-Rojas for experimental support, and Neil Mathur for discussions. SKN is grateful for support from the Royal Society through a Dorothy Hodgkin Fellowship. VN acknowledges the Herchel Smith Fund, University of Cambridge for a Fellowship. This work was supported by the EPSRC Cambridge NanoDTC, EP/G037221/1.

Received:
Revised:
Published online:

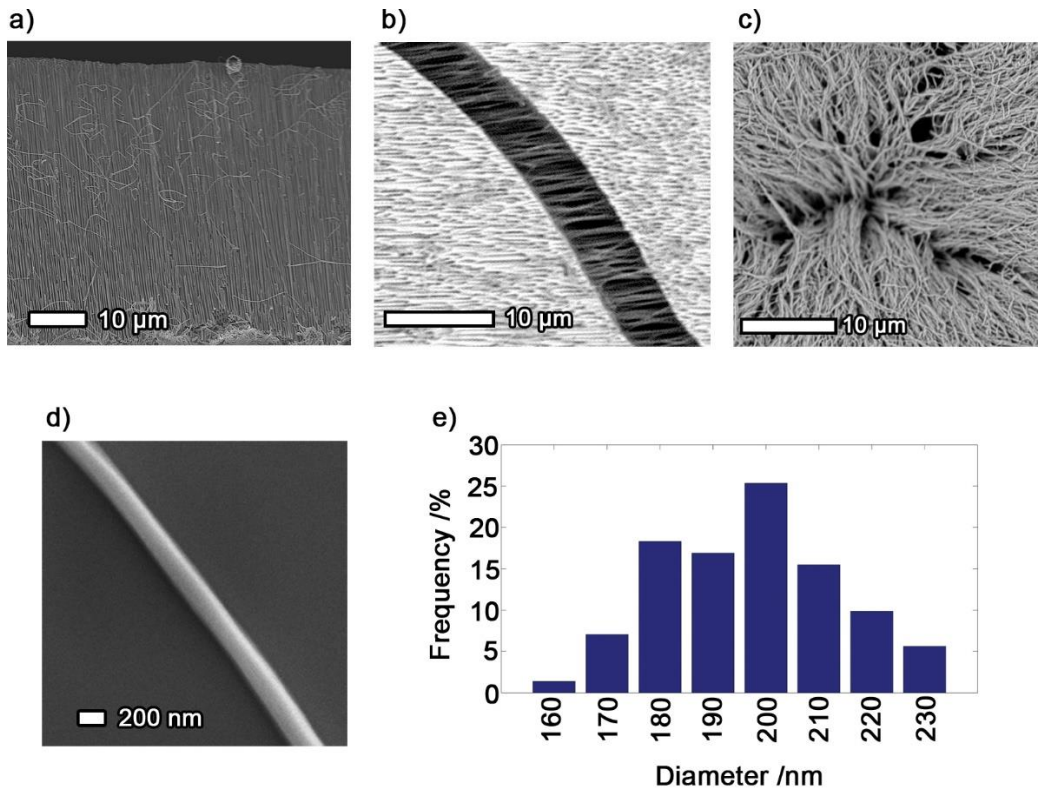

**Figure 1.** Template-grown P(VDF-TrFE) nanowires. (a) Aligned nanowires in AAO template viewed within a cleaved cross-section. Some nanowires appear to have been pulled out of the pores during the cleaving process. (b) Nanowires observed within a fracture when viewing the cleaved edge of a nanowire-filled template. (c) A partially dissolved nanowire-filled template viewed from above, revealing the forest of nanowires within. (d) A single nanowire drop cast onto a silicon substrate after being dispersed in solution by completely dissolving the template. (e) Diameter distribution of the nanowires.

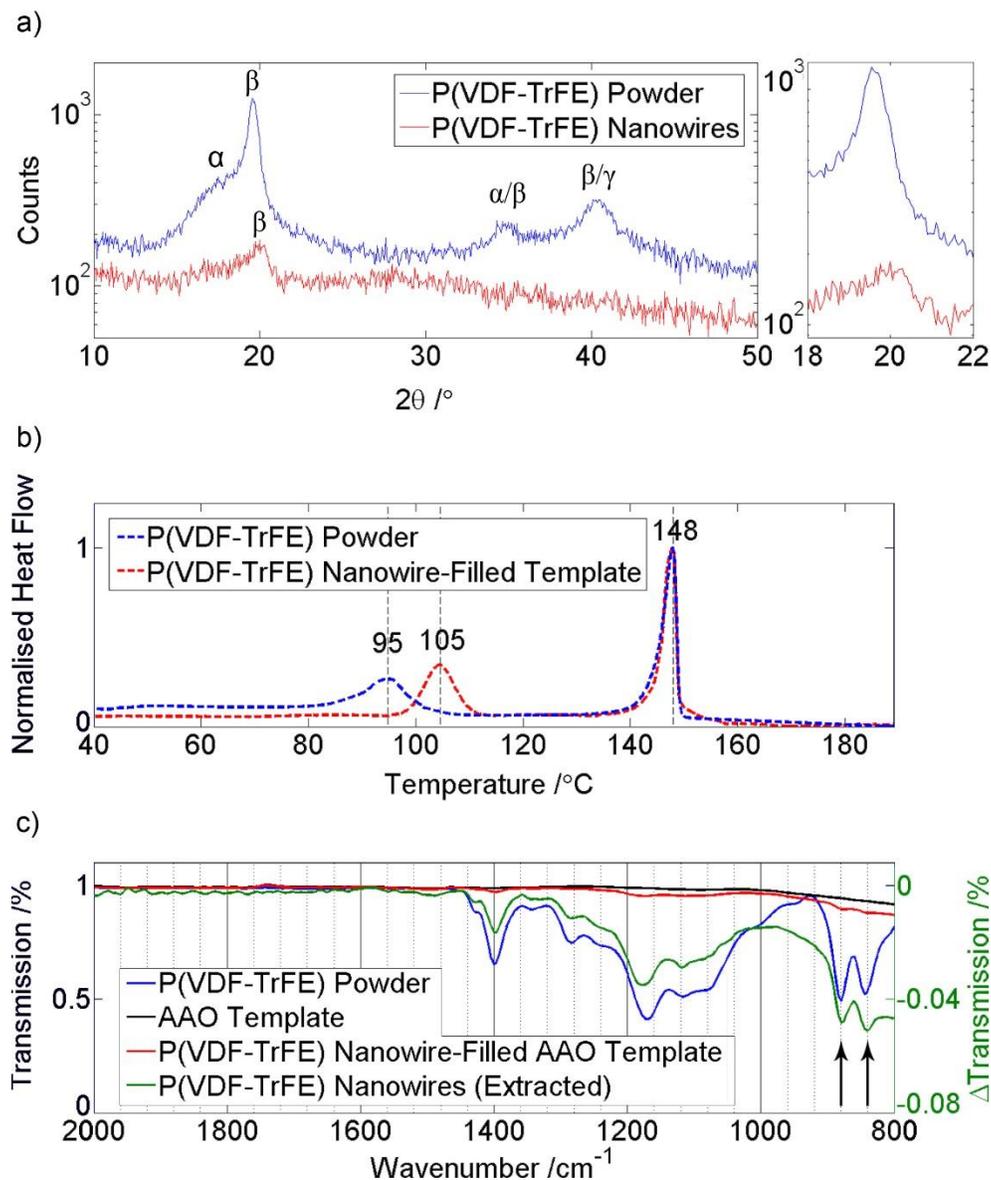

**Figure 2.** Characterization of structure and morphology of template-grown nanowires. (a) XRD patterns from P(VDF-TrFE) nanowires and P(VDF-TrFE) powder of the same composition. The peak at $2\theta = 19.7°$ corresponding to diffraction in (200) and (110) planes represents the polar

β-phase, and is dominant in the nanowire XRD spectrum. The XRD pattern of the powder shows a mixture of α, β and γ phases in comparison. The right panel is a focus graph of the XRD patterns between 18° and 22°. (b) Heat flow as a function of temperature during DSC scans of P(VDF-TrFE) powder and of a ground P(VDF-TrFE) nanowire-filled AAO template. The curves have been normalized by the melting peaks to facilitate comparison of the results. (b) ATR-FTIR spectrograms of P(VDF-TrFE) powder, a ground AAO empty template and a P(VDF-TrFE) nanowire-filled AAO template. The IR data for the nanowires (green line) is extracted by subtracting the empty template transmission data from the filled template transmission data and is plotted against the vertical scale on the right. The arrows indicate the positions of the trough at ∼880 cm$^{-1}$ which is associated with trans links in the carbon chains and the trough at ∼840 cm$^{-1}$ which is associated with more than three consecutive trans links, the latter therefore being associated with the β-phase.

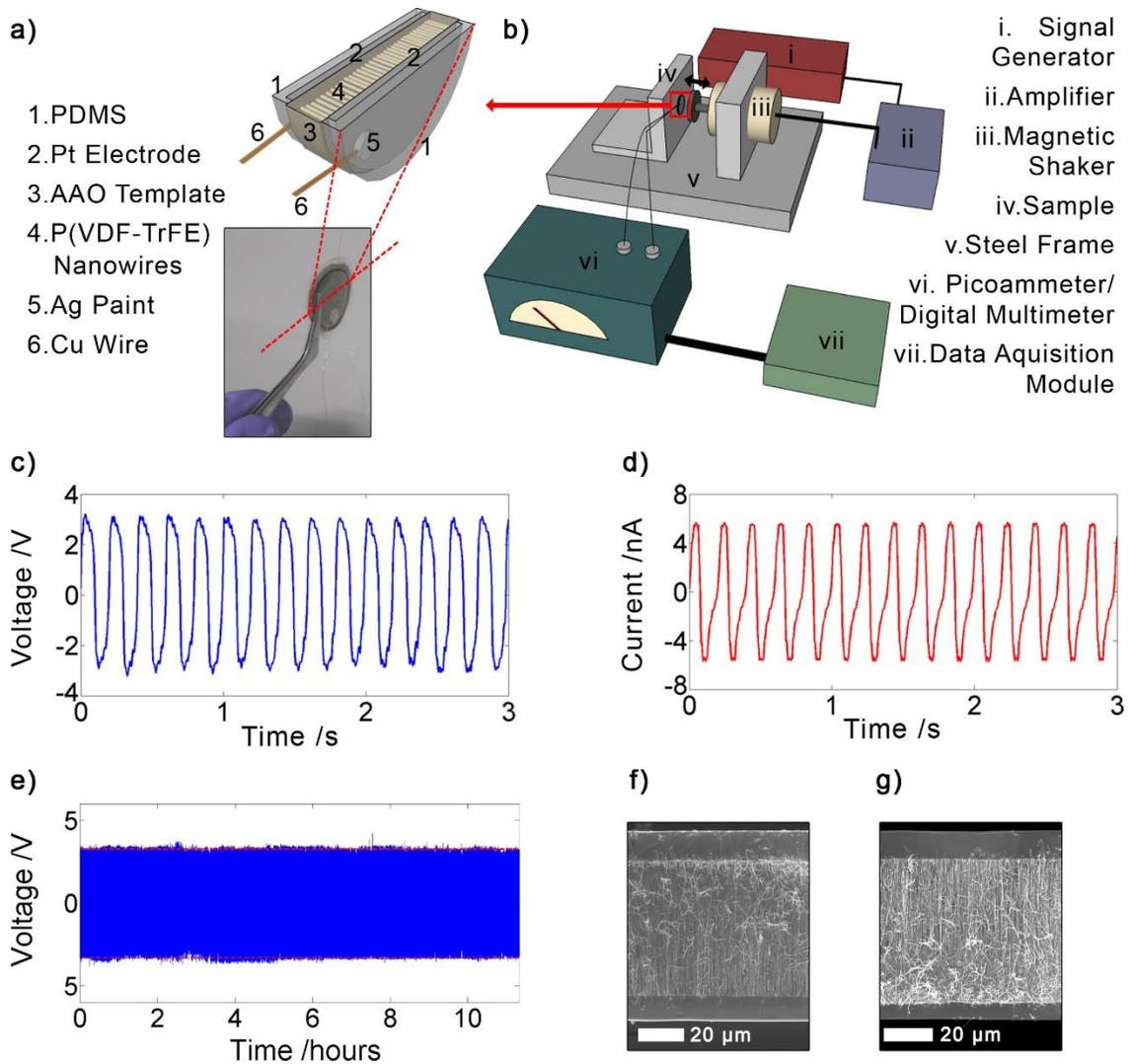

**Figure 3.** Energy harvesting measurements and results. (a) Schematic of the template-grown P(VDF-TrFE) nanowire-based nanogenerator layout with a picture of an actual device for visualization. (b) Schematic of the experimental set-up used in the piezoelectric energy harvesting measurements. (c) Typical output voltage and (d) output current characteristics of the nanogenerator. (e) Output voltage recorded as a function of time during fatigue testing over 12 hours. (The horizontal red dotted lines serve as guides to the eye). (f) Cross-sectional SEM image of a nanogenerator taken before and (g) after fatigue testing. Some nanowires appear to have been pulled out of the pores during the cleaving process.

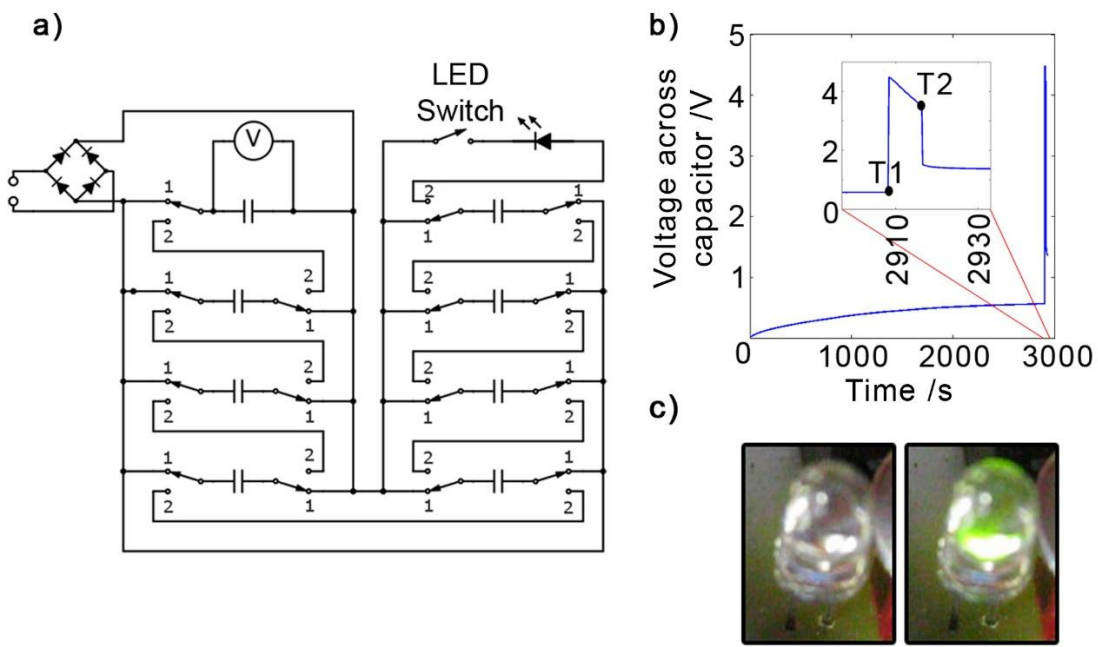

**Figure 4.** Lighting an LED using a template-grown nanogenerator. (a) Diagram of the circuit used for storing the power generated from a nanogenerator before releasing it to light an LED. All of the capacitors shown have a capacitance of 22 μF each. (b) Voltage across one of the capacitors in the circuit shown in (a) while charging from a nanogenerator and subsequent discharge across an LED. The inset shows a zoomed-in view around the region of voltage step-up (T1) and discharge (T2). (c) Photographs showing the LED going from unlit to lit during the discharge.

**Supporting Information**

# A Scalable Nanogenerator based on Self-Poled Piezoelectric Polymer Nanowires with High Energy Conversion Efficiency

R.A. Whiter, V. Narayan and S. Kar-Narayan

## S1. Polymer infiltration of nanoporous AAO template using silver film

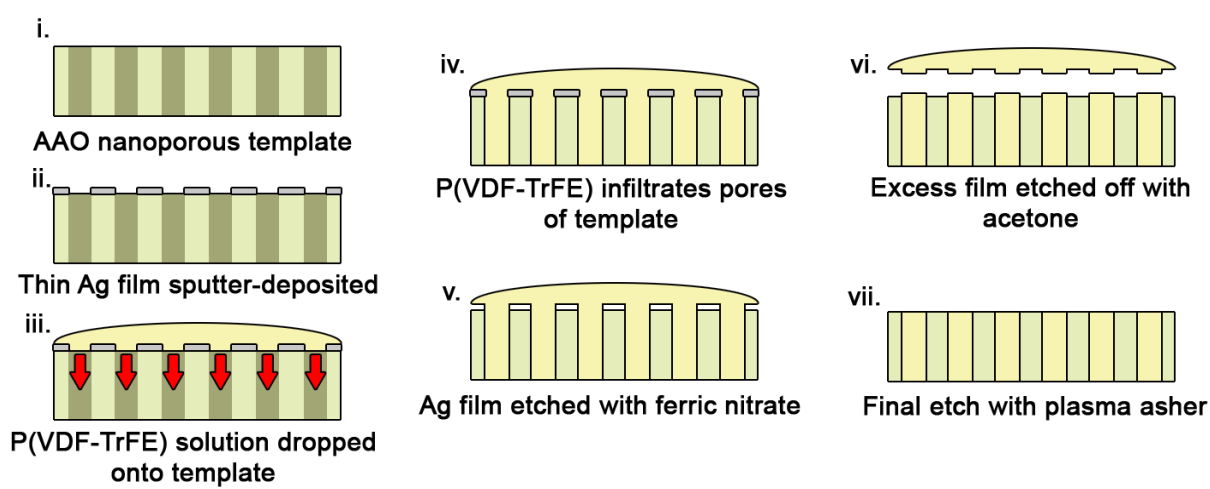

**Supplementary Figure 1**

Prior to deposition of the polymer, the templates were coated with a thin layer of silver ~15 nm thick using a desktop sputter coater (*Emitech k550*) (Step i-ii). The film was thin enough not to cover the holes of the pores but instead form around them. Following infiltration of the P(VDF-TrFE) solution (Steps iii-iv) the filled templates were then first placed in an aqueous solution of ferric nitrate of concentration 1 M which dissolved the silver layer over a period of ~1 hour without reacting with the polymer or templates (Step v). The templates were then rapidly transferred to a beaker of water to remove the ferric nitrate and then rapidly transferred to a beaker of acetone within which they were sonicated for 2 minutes (Step vi). In the final step the templates were plasma ashed for 1 minute using a microwave plasma reactor (*MRC200, Cambridge Fluid Systems*) (Step v). This methodology resulted in minimal etching of the nanowires while ensuring removal of the film.

## S2. Throughput and cost of the template-wetting method

Our processing time for P(VDF-TrFE) nanowires in AAO templates was ~ 6 hours. Several nanowire-filled templates can be processed in parallel. The mass of nanowires grown per template is ~17 mg (equivalent to $10^{10}$ fibers). Therefore the yield is 17/6 = 2.8 mg (equivalent to ~ $2 \times 10^8$ fibers) per hour per template.

In terms of cost, the cost per template of polymer and solvent are ~£0.04 and ~£0.003 respectively, while the cost of aluminium is ~£1 per kg. For the dimensions of the template we use, this gives an approximate overall cost of ~£2.61 per gram of P(VDF-TrFE) nanowires. We expect this cost to be significantly reduced during large-scale production where the raw materials would be cheaper.

### S3. Dissolving template to release nanowires into solution

To release the nanowires into solution the templates were submerged in vials with a minimal amount (~1 ml) of a 10% wt aqueous phosphoric acid solution and sonicated for ~1 hour and then left submerged for ~24 hours to ensure complete dissolution of the template. The vials were then topped up with deionised water and placed in a centrifuge (*Sigma 1-14 Microfuge*) for 1 hour at 14,000 RPM. This resulted in the nanowires collecting on the bottom of the vial and allowed for the phosphoric acid solution to be removed and replaced with water within which the nanowires were re-suspended. This process was repeated several times to wash out all of the phosphoric acid. From the water solutions the nanowires could then be drop-cast onto substrates.

### S4. Energy Harvesting Set-Up

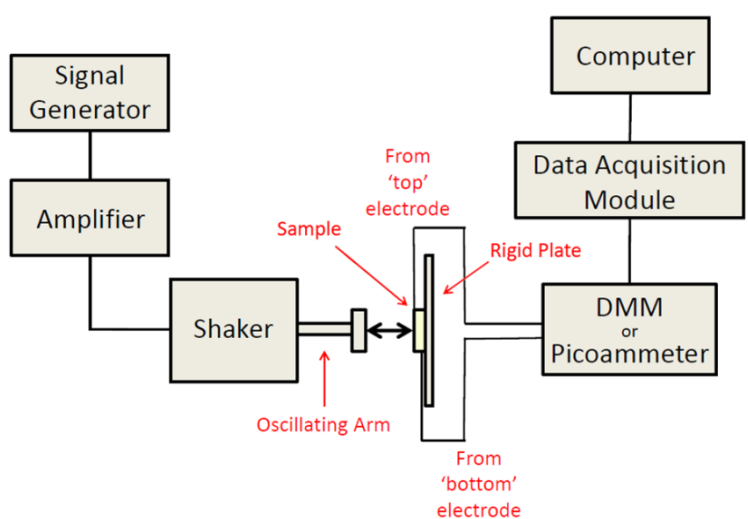
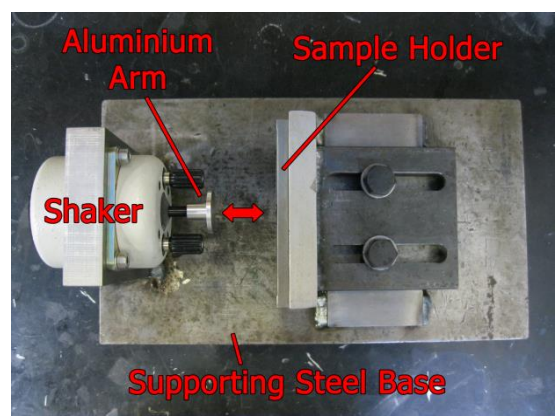

Supplementary Figure 2b

**Supplementary Figure 2**

Signal Generator - TG1304, Thurlby Thandor
Amplifier - PA25E-CE, LDS Systems
Permanent Magnetic Shaker - V100, LDS Systems
Digital Multimeter – KUSB-3116, Keithley
Picoammeter – 6487, Keithley
Data Acquisition Module - KUSB-3116, Keithley

### S5. Strain rate

The strain, $\varepsilon$, in the nanowires can be calculated using $\varepsilon = \sigma/E$ with $\sigma = V_p/g_{33}L$ where $V_p$ is the peak output voltage and $g_{33}$ is the piezoelectric coefficient. Taking $g_{33} = 0.43$ Vm N$^{-1}$, $E =$ 0.25 GPa,[2] $L = 60$ μm and $V_p = 3$ V gives the strain as $4.65 \times 10^{-6}$. Considering that our device underwent compressive strain for a quarter of the oscillation period, the strain rate is taken to be $4f\varepsilon \approx 0.01\%$ s$^{-1}$.

## S6. Energy Efficiency Calculation

The electrical energy output from a device per cycle can be calculated as $\int IV dt$ where *I* and *V* are the output current and voltage respectively measured at time *t* and the integral is over one oscillation. This was found to be $2.05 \times 10^{-9}$ J when integrating over 20 cycles and then averaging. Supplementary Figure 3 shows a plot of *I*V* for 3 cycles with the area according to one cycle shaded.

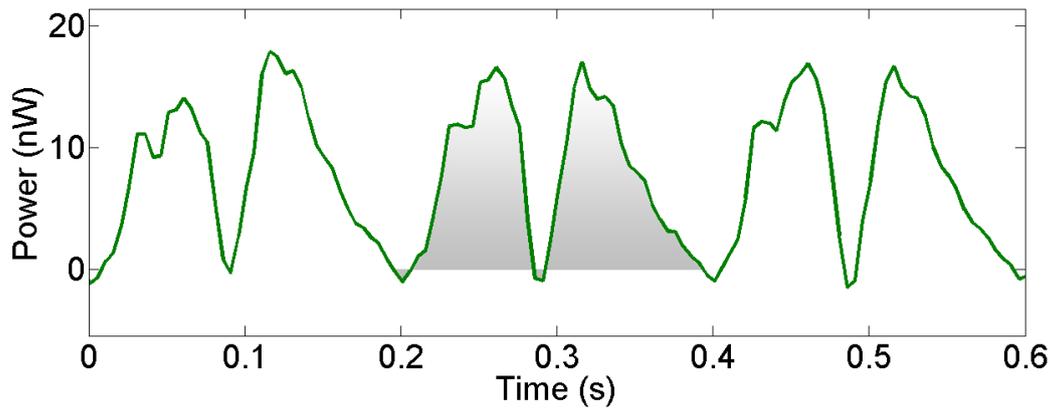

**Supplementary Figure 3**

The strain energy may be approximated as $\frac{1}{2}AL\varepsilon^2 E$ [1] where *A* is the combined cross-sectional area of the nanowires within the device, $L = 60$ μm is the original length of the nanowires, $\varepsilon$ is the strain in the nanowires after deformation and *E* is the Young's Modulus (see S5 above). The strain energy is then calculated to be $1.84 \times 10^{-8}$ J.

The energy conversion efficiency taken to be the calculated electrical energy divided by the calculated strain energy is found to be 11.2%.

## S7. Multiple devices in series

A stack of 3 devices wired in series was achieved by taping the devices down on top of each other in the energy harvesting set-up described above. The wires from the electrodes of the middle device were then connected to the respective adjacent electrodes of the two outer devices and the electrodes at either end of the stack were wired to the circuit shown in Figure 4a of the main text. The stack was then impacted in the same manner as for the single devices. Using a frequency of 20 Hz and amplitude of 1 mm for the oscillating arm, the stack was shown to charge the capacitors such that the LED could be lit in 703 s as shown in Supplementary Figure 4. To charge the capacitors such that they had the same voltage across them with a single device took 2908 s as shown in Figure 4b of the main text.

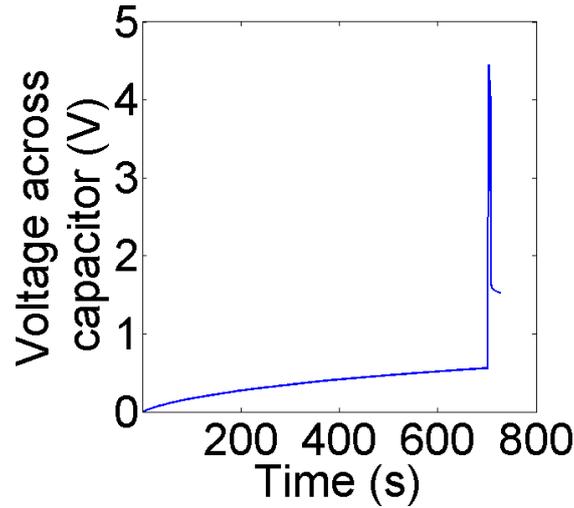

**Supplementary Figure 4**

## S8. Overall efficiency of the device

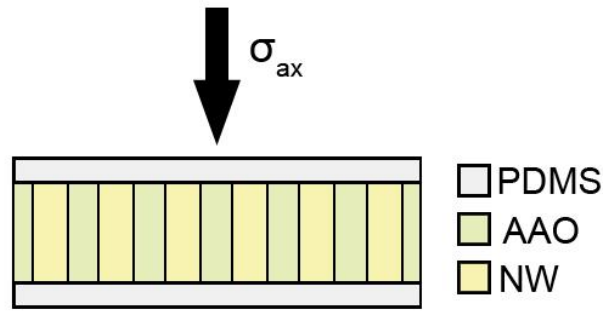

**Supplementary Figure 5**

The strain in the nanowires was found to be $4.65 \times 10^{-6}$ (see S5 above). Since the nanowires are embedded in the AAO template, this is the same strain in the AAO template. The individual stresses in the nanowires and the template are given as:

$\sigma_{NW} = \varepsilon_{NW} E_{NW}$ and $\sigma_{AAO} = \varepsilon_{AAO} E_{AAO}$, where $\varepsilon_{NW} = \varepsilon_{AAO} = \varepsilon_{NW/AAO}$

The resultant axial compressive stress, $\sigma_{ax}$, on the nanowire-filled template is given by:
$\sigma_{ax} = f(\sigma_{NW}) + (1-f)\sigma_{AAO}$, where $f$ is the volume fraction of the nanowires and is equal to the porosity of the template, i.e. 0.5 in this case.

The equivalent Young's modulus of the nanowire-filled template (by the rule of mixtures) is:
$E_{NW+AAO} = f(E_{NW}) + (1-f) E_{AAO}$

Taking $E_{AAO} = 15$ GPa [3], we find that $E_{NW+AAO} = 7.6$ GPa and $\sigma_{ax} = 4.1 \times 10^4$ Pa. This is the same stress generated in the PDMS layers (see Supplementary Figure 5), and is equal to the total stress generated by the impacting arm. Given that the device area is ~ 2 cm$^2$, this translates to an impacting force, $F$, of about 9 N.

The strain in the PDMS layer is thus $\varepsilon_{PDMS} = \sigma_{ax} / E_{PDMS} = 0.008$, taking $E_{PDMS} = 1.5$ MPa. [4]

The overall strain in the nanogenerator is then given by:
$\varepsilon_{NG} = f'(\varepsilon_{PDMS}) + (1-f')\varepsilon_{NW/AAO} = 0.008$, where $f'$ is the volume fraction of PDMS in the nanogenerator, i.e. 0.3 in this case since thickness of PDMS layer = 10 μm. The strain is predominantly in the PDMS layer due to its relatively low Young's Modulus as compared to the nanowire-filled template.

The mechanical energy transferred to the nanogenerator by the impacting arm per cycle is $F \Delta L = 1.44$ μJ, where $\Delta L$ is the stopping distance of the arm as it impacts the nanogenerator, where $\varepsilon_{NG} = \Delta L / L$.

The total electrical energy harvested by the nanogenerator and subsequently stored in the bank of 8 capacitors after being impacted at 20 Hz for 1000 s (i.e. 20000 impacting cycles) is $0.5(CV^2) \times 8 = 44$ μJ, where $C = 22$ μF is the capacitance of each capacitor, and $V = 0.5$ V after 1000 s. The total mechanical energy required to generate this energy is $20000 \times 1.44$ μJ $= 0.028$ J.[5]

Therefore the overall device efficiency is the total electrical energy harvested divided by the total mechanical energy required. Under our experimental conditions, the overall efficiency of the nanogenerator we tested is ~ 0.2 %.

**S9. Control energy harvesting experiment with empty AAO template**

An empty AAO template, i.e. devoid of piezoelectric nanowires, was electroded, wired up, encapsulated in PDMS and tested under the same conditions as the nanogenerator in Figure 3 of the manuscript. There was virtually negligible voltage and current output from this control sample as can be seen from Supplementary Figure 6, showing that the P(VDF-TrFE) nanowires form the only piezo-active part of the nanogenerator.

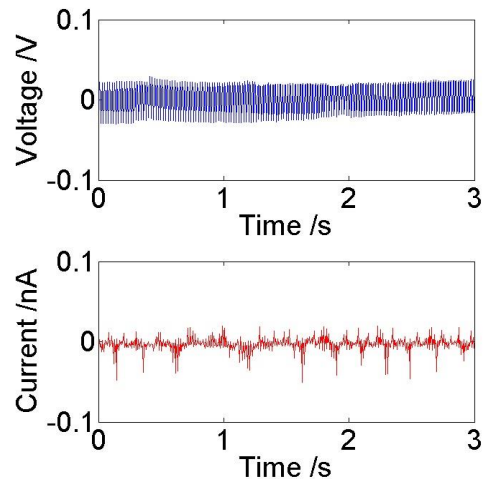

**Supplementary Figure 6**